\documentclass[11pt]{article}
\usepackage{graphicx}

\begin{document}

\def\xslash#1{{\rlap{$#1$}/}}
\def \p {\partial}
\def \dd {\psi_{u\bar dg}}
\def \ddp {\psi_{u\bar dgg}}
\def \pq {\psi_{u\bar d\bar uu}}
\def \jpsi {J/\psi}
\def \psip {\psi^\prime}
\def \to {\rightarrow}
\def\bfsig{\mbox{\boldmath$\sigma$}}
\def\DT{\mbox{\boldmath$\Delta_T $}}
\def\xit{\mbox{\boldmath$\xi_\perp $}}
\def \jpsi {J/\psi}
\def\bfej{\mbox{\boldmath$\varepsilon$}}
\def \t {\tilde}
\def\epn {\varepsilon}
\def \up {\uparrow}
\def \dn {\downarrow}
\def \da {\dagger}
\def \pn3 {\phi_{u\bar d g}}

\def \p4n {\phi_{u\bar d gg}}

\def \bx {\bar x}
\def \by {\bar y}

%  Start to revise for addressing referee's comments in the second round  17.02.2017. 

\begin{center}
{\Large\bf  Light-Cone Singularities and Transverse-Momentum-Dependent Factorization at Twist-3 }
\par\vskip20pt
A.P. Chen$^{1,2}$ and J.P. Ma$^{1,2,3}$     \\
{\small {\it
$^1$ Institute of Theoretical Physics, Chinese Academy of Sciences,
P.O. Box 2735,
Beijing 100190, China\\
$^2$ School of Physical Sciences, University of Chinese Academy of Sciences, Beijing 100049, China\\
$^3$ Center for High-Energy Physics, Peking University, Beijing 100871, China  
}} \\
\end{center}
\vskip 1cm
\begin{abstract}
We study transverse-momentum-dependent factorization at twist-3 for Drell-Yan processes. The factorization 
can be derived straightforwardly at leading order of $\alpha_s$. But at this order we find that 
light-cone singularities already exist and effects of soft gluons are not correctly factorized. 
We regularize the singularities with gauge links off the light-cone and introduce a soft factor to factorize the effects of soft gluons. Interestingly, the soft factor must be included in the definition 
of subtracted TMD parton distributions to correctly 
factorize the effects of soft gluons. We derive the Collins-Soper equation for one of twist-3 TMD parton distributions. The equation can be useful for resummation of large logarithms terms appearing in the corresponding structure function in collinear factorization. 
However, the derived equation is nonhomogeneous. This will make the resummation complicated.  
\vskip 5mm
\noindent
% PACS numbers
\end{abstract}
\vskip 1cm
\par 

\par
Differential cross-sections of hadron collisions with observed small transverse momenta in final states, 
like Drell-Yan processes with the lepton pair at low transverse momentum $q_\perp$, are sensitive to the transverse motion of partons inside hadrons. To consistently describe these processes in QCD, one needs to establish their Transverse-Momentum-Dependent(TMD) factorization, in which nonperturbative-
and perturbative effects are systematically separated.   
Using the proven factorizations one can extract from experiment various TMD parton distributions which contain information about inner structure of hadrons. 
\par 
With TMD factorization  the perturbative effects
due to large energy scale $Q$ in processes can be safely calculated with perturbation theory, and the 
nonperturbative effects are described with TMD parton distributions defined with QCD operators.
To perform such a factorization one expands differential cross-section 
in the inverse power of $Q$. At the leading power of $1/Q$ or twist-2,  
TMD factorization has been established for 
several processes like $e^+e^-$-annihilations\cite{CS},  Drell-Yan processes\cite{CSS,JMYP} and 
Semi-Inclusive Deeply Inelastic Scattering(SIDIS)\cite{JMY,CAM}. But, 
 TMD factorization at next-to-leading power or twist-3 has been not studied at the rigorous level as that of factorizations at twist-2, although TMD factorization at twist-3  for some observables
can be derived in a straightforward way at leading order of $\alpha_s$. Some results about this can be found, 
e.g., in \cite{BDGMMS,BMP,BJM,PMT,ABMP,LS,AMS}. In this letter, we take unpolarized Drell-Yan process as an 
example to examine TMD factorization at twist-3. 
\par  
It is well-known that light-cone singularities appear in twist-2 TMD parton distributions, if they are   
defined with light-cone gauge links as used in the definitions of parton distribution functions in collinear factorizations. These singularities appear at the next-to-leading order of $\alpha_s$ and can be regularized 
by using gauge links off light-cone direction. 
In general one can expect that twist-3 TMD parton distributions defined with light-cone gauge links will also have light-cone singularities as shown in \cite{GHMS}.  
With explicit calculations we will show that light-cone singularities already appear at leading order of $\alpha_s$ in twist-3 TMD 
factorization. We regularize them with off light-cone gauge links. 
Our result shows that a soft factor is needed even at the leading order of $\alpha_s$ to 
correctly factorize soft-gluon contributions.  
However, unlike the case of TMD factorization at twist-2, the soft factor has to be implemented in twist-3 factorization in an unique way from our result.     
\par 
The regularization of the light-cone singularities introduces a regulator-dependence in TMD parton distributions. The dependence is governed by Collins-Soper equations. It is interesting to note that this type of equations can be used to resum  large log's of $q_\perp/Q$ in perturbative coefficient functions in collinear factorization, as shown in \cite{CS,CSS,JMY}. The behavior of all structure functions of unpolarized Drell-Yan processes in the limit $q_\perp \to 0$ has been studied in \cite{BQR,BV}, where one finds that 
it can be difficult to resum large log's in one of the structure functions. It is this structure which is relevant to the twist-3 TMD factorization studied here. We will derive the Collins-Soper equation 
for the twist-3 TMD parton distribution involved here. Our result shows that the resummation can be difficult 
because the needed Collins-Soper equation is not homogeneous.       
  
\par 
We consider the unpolarized Drell-Yan process:
\begin{equation}
  h_A ( P_A) + h_B(P_B) \to\gamma^* (q) +X \to \ell^- ( k_1) + \ell ^+ (k_2) + X,
\end{equation}
where initial hadrons are unpolarized.
We will use the  light-cone coordinate system. In this system a
vector $a^\mu$ is expressed as $a^\mu = (a^+, a^-, \vec a_\perp) =
((a^0+a^3)/\sqrt{2}, (a^0-a^3)/\sqrt{2}, a^1, a^2)$ and $a_\perp^2
=(a^1)^2+(a^2)^2$. We introduce two light-cone vectors: $n^\mu =(0,1,0,0)$ and $l^\mu = (1,0,0,0)$, and
two transverse tensors $g_\perp^{\mu\nu} = g^{\mu\nu} - n^\mu l^\nu - n^\nu l^\mu$, and $\epsilon_\perp^{\mu\nu}= \epsilon^{\alpha\beta\mu\nu} l_\alpha n_\beta$. 
In this system, the momenta of initial hadrons are given as: 
\begin{equation}
P_A^\mu \approx (P_A^+, 0, 0,0),  \ \ \ \  P_B^\mu \approx (0, P^-_B, 0,0).
\end{equation}
The angular distribution is characterized by four structure functions\cite{LT}: 
\begin{eqnarray} 
\frac{ d\sigma}{d^4 q d \Omega} = \frac{\alpha^2_{em}}{2(2\pi)^4 S^2 Q^2} \biggr [ 
  W_T (1+\cos^2\theta)   + W_\Delta  \sin(2\theta) \cos\phi + W_L (1-\cos^2\theta)
   +W_{\Delta\Delta} \sin^2\theta \cos(2\phi) \biggr ]
\label{AD}    
\end{eqnarray} 
where $\Omega$ is the solid angle of the observed lepton in Collins-Soper frame.   
In collinear factorization one finds in the limit $q_\perp\to 0$ that the first two structure functions have power divergences, $Q^2/q_\perp^2$ and $Q/q_\perp$, respectively\cite{BQR,BV}. The structure function $W_\Delta$ is relevant to our study here. It is noted that the resummation of large log terms 
in the perturbative coefficient function of $W_T$ has been studied extensively. But it is not clear how to resum the large log terms in $W_{\Delta}$, $W_L$ and $W_{\Delta\Delta}$, as discussed in \cite{BQR,BV}. 
  
\par

The hadronic tensor for the process is defined as:
\begin{eqnarray}
W^{\mu\nu} = \sum_X \int \frac{d^4 x }{(2\pi)^4} e^{iq \cdot x} \langle h_A (P_A) h_B(P_B)  \vert
    \bar q(0) \gamma^\nu q(0) \vert X\rangle \langle X \vert \bar q(x) \gamma^\mu q(x) \vert
     h_B(P_B)h_A (P_A)  \rangle
\end{eqnarray}
where spins of initial hadrons are averaged. For brevity we will take the electric charge of quarks 
as $1$ in this work. We are interested in the kinematical region 
of $q^2_\perp \ll Q^2$. In this region the tensor can be decomposed as\cite{BQR}
\begin{equation}
W^{\mu\nu} = - g_\perp^{\mu\nu} W_\perp 
              +\frac{1}{q^-} \left ( q_\perp^\mu l^\nu +  q_\perp^\nu l^\mu \right ) W_l
              +\frac{1}{q^+} \left ( q_\perp^\mu n^\nu +  q_\perp^\nu n^\mu \right ) W_n
               + \cdots, 
\label{WSQT}                
\end{equation}
where $\cdots$ stand for terms whose effects are power-suppressed. We assume that the polarization of the 
leptons in the final state is not observed. Then we need only to consider the symmetric part of $W^{\mu\nu}$.   
In the small $q_\perp$ region, the above structure functions are related to $W_T$ and $W_\Delta$ as:
\begin{equation} 
 W_T = W_\perp, \quad\quad  W_\Delta = \frac{q_\perp}{Q} \left ( W_n -W_l\right ). 
\end{equation} 
For $W_\perp$ one can derive its twist-2 factorization\cite{CSS}, while for $W_{l,n}$ TMD factorization at twist-3 is needed.   
 
\par 
TMD parton distributions are defined as hadronic matrix elements of QCD operators. As mentioned before, 
we will use the gauge link off the light-cone. This will regularize light-cone singularities as shown later. 
We introduce: 
\begin{eqnarray}
{\mathcal L}_u (\xi) = P \exp \left ( -i g_s \int_{-\infty}^0  d\lambda
     u\cdot G (\lambda u + \xi) \right ) ,
\end{eqnarray} 
with $u^\mu=(u^+,u^-,0,0)$ and $u^-\gg u^+$. The TMD quark distributions of $h_A$ are defined with the quark density matrix  
\begin{eqnarray}
{\mathcal M}_{ij} (x,k_\perp)  =  \int \frac{ d\xi^-d^2\xi_\perp } {(2\pi)^3} e^{ -i x \xi^- P_A^+ - i \xi_\perp \cdot k_\perp}
\langle h_A  \vert  \biggr ( \bar q(\xi) {\mathcal L}_u (\xi) \biggr )_j \biggr ({\mathcal L}_u^\dagger (0)
  q (0) \biggr )_i \vert h_A  \rangle \biggr\vert_{\xi^+ =0} ,
\label{DENMG}
\end{eqnarray}
where $i$ and $j$ are color- and Dirac-spinor indices.
Similarly one has the quark density matrix of $h_B$, where the gauge link ${\mathcal L}_v$ is along the direction $v^\mu=(v^+,v^-,0,0)$ and $v^+\gg v^-$.  
We consider that the hadrons are unpolarized. In this letter, we work with Feynman gauges which is a non-singular gauge. In singular gauges transverse gauge links at $\xi^-=\infty$ should be added to make 
the density matrix gauge invariant\cite{TMDGL,TMDGL1}. In the framework of Soft-Collinear Effective Theory(SCET)
such transverse gauge links are also needed for gauge invariance as shown in \cite{IC1}.      
The parameterization of the density matrix have been studied in \cite{PMT,ABMP,BoMu,GMS}, where twist-2- and twist-3 TMD quark distributions are defined.   
The density matrix for the unpolarized hadron is parameterized with TMD parton distributions up to twist-3 as: 
\begin{eqnarray}
{\mathcal M}_{ij} (x,k_\perp )  &=& \frac{1}{2N_c} \biggr \{  f_1(x,k_\perp)\gamma^- +h_{1}^\perp(x,k_\perp)\sigma^{\mu-}\frac { k_{\perp\mu} }{M_A}
+ \frac{M_A}{P_A^+}\biggr [  e (x,k_\perp) + \frac{1}{M_A} f^\perp (x,k_\perp) \gamma\cdot k_\perp 
\nonumber\\     
    && +  h(x,k_\perp) \sigma^{-+} 
  - \frac{1}{M_A} g^{\perp} (x, k_\perp) \epsilon_\perp^{\mu\nu} \gamma_\mu \gamma_5  k_{\perp\nu}    
\biggr ] \biggr\}_{ij} +\cdots, 
\label{DECMB}
\end{eqnarray} 
where $\cdots$ stand for those of higher twist. In Eq.(\ref{DECMB}) the first two terms are of twist-2.  
$f_1$, $f^\perp$ and $g^\perp$ are defined with chirality-even operators, while 
$e$, $h$ and $h_1^\perp$ are defined with chirality-odd operators. 
In this letter we will only consider the contributions to $W^{\mu\nu}$ involving chirality-even operators. 
The contributions involving chirality-odd operators are irrelevant to the behavior the small-$q_\perp$ or to the large log's of $q_\perp/Q$  
appearing in the collinear factorization. They 
will be studied in a separate work.
The defined distributions depend not only on the momentum fraction $x$ and the transverse momentum $k_\perp$, 
but also on the renormalization scale $\mu$ and the parameter $\zeta_u$ defined as $\zeta_u^2 = ( 2u\cdot P_A)^2/u^2$. Similarly one can define TMD antiquark distributions of $h_B$. The parameter $\zeta_v$ related to $h_B$ is then defined 
as $\zeta_v^2 = (2v \cdot P_B)^2/v^2$. In Eq.(\ref{DECMB}) it is implied that  $\zeta_u$ is large but finite and any contribution proportional to power of $\zeta_u^{-1}$ is neglected. 
\par 

\begin{figure}[hbt]
\begin{center}
\includegraphics[width=14cm]{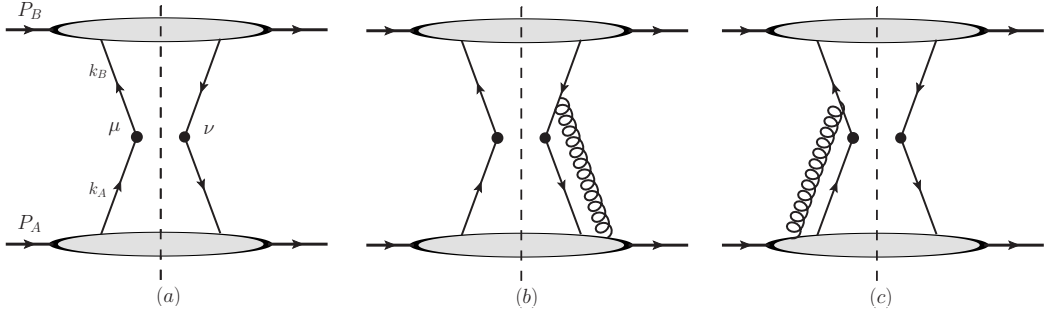}
\end{center}
\caption{The Feynman diagrams for $W^{\mu\nu}$. The black dots stand for the insertion of the electromagnetic current operator.    } 
\label{Feynman-dg2}
\end{figure}
\par

\par 
TMD factorization at leading order of $\alpha_s$  can be derived straightforwardly by the diagram expansion. 
The factorized form for $W_l$ and $W_n$ have been essentially derived in \cite{PMT}. 
If we consider the case that the process is initiated by a quark from $h_A$ and an antiquark from $h_B$, 
then one needs to consider  up to twist-3 diagrams given in Fig.1. In Fig.1 the bubbles are jet-like Green functions related to the initial hadrons. From Fig.1a one can obtain not only twist-2 contributions to $W_\perp$, but also twist-3 contributions to $W_l$ and $W_n$. At the considered order of $Q^{-1}$ the bubbles or jet-like Green functions in Fig.1a are essentially the quark density matrices like the one defined in Eq.(\ref{DENMG}).  We have:  
\begin{eqnarray} 
W^{+\nu}\biggr\vert_{1a} = W^{\nu +}\biggr\vert_{1a} = \frac{q_\perp^\nu}{N_c P_B^-} \int d^2 k_{A\perp} d^2 k_{B\perp} \frac{ \vec k_{B\perp}\cdot \vec q_\perp}{q_\perp^2} 
    \delta^2 (k_{A\perp}+k_{B\perp}-q_\perp)
  f_1 (x,k_{A\perp}) \bar f^\perp (y,k_{B\perp})  , 
\label{FAC1}    
\end{eqnarray}
where $\bar f^\perp$ are TMD antiquark distributions of $h_B$. The $+$- or $-$-component of $q$ are given by 
$q^+ =x P_A^+$ or  $q^- = y P_B^-$, respectively.  
\par 
The calculation of Fig.1b or Fig.1c is straightforward. From Fig.1c we have the result in the first step:  
\begin{eqnarray} 
W^{+\nu}\biggr\vert_{1c}  =  \frac{1}{N_c P_B^-} \int d^2 k_{A\perp} d^2 k_{B\perp}  \delta^2 (k_{A\perp} +k_{B\perp} -q_\perp) 
      \bar f_1 (y, k_{B\perp} ) \phi_3^\nu (x,k_{A\perp}),
\label{W3G}           
\end{eqnarray}
where $\phi_3^\nu$ is the quark-gluon correlator defined with $k_A^\mu =(xP_A^+,0,\vec k_{A\perp})$ as:  
\begin{eqnarray}
 \phi_3^\nu (x,k_{A\perp}) = \frac{g_s}{2} \int_{-\infty}^{0}  d\lambda   \int\frac{d\xi^-d^2\xi_\perp}{(2\pi)^3} e^{- i \xi\cdot k_A } 
\langle h_A \vert \bar \psi( \xi  ) \gamma^+ \left ( g_\perp^{\nu\rho} -i\epsilon_\perp^{\nu\rho} 
    \gamma_5 \right )  G^{+}_{\ \  \rho} (\lambda n  ) \psi ( 0 ) \vert h_A \rangle \biggr\vert_{\xi^+=0}. 
\label{PHI3} 
\end{eqnarray}
In Fig.1c only one-gluon exchange has been shown. Additional exchanges of collinear gluons also give 
contributions at the same power. These contributions can be summed into gauge links which 
appear between field operators in the above. We have suppressed these gauge links. 
The final result for the symmetric part of $W^{+\nu}$ is the sum of Fig.1a and Fig.1c.  
Later, we will show that in fact $\phi_3^\nu$ has the light-cone singularity. 
The result in Eq.(\ref{W3G}) can be expressed with TMD quark distributions.  By using equation 
of motion, as shown in \cite{PMT},  
the defined quark-gluon correlator can be expressed with TMD quark distributions:  
\begin{equation} 
 \phi_3^\nu (x, k_\perp) = k_\perp^\nu f_1 (x, k_\perp) - x k_\perp^\nu f^\perp (x, k_\perp) + i k_\perp^\nu g^\perp (x,k_\perp).
\label{RE}         
\end{equation} 
It should be noted that the TMD quark distributions in the right-hand side are not exactly those 
defined in Eq.(\ref{DECMB}). They are defined with gauge links along the direction 
$n^\mu$. Replacing them with those defined in Eq.(\ref{DECMB}) and using the relation 
we obtain the final result of $W^{+\nu}$:
\begin{eqnarray} 
W^{+\nu}  &=& \frac{q_\perp^\nu}{N_c P_B^-} \int d^2 k_{A\perp} d^2 k_{B\perp} \delta^2 (k_{A\perp}+k_{B\perp}-q_\perp)
  \biggr [  \frac{ \vec k_{B\perp}\cdot \vec q_\perp}{q_\perp^2} f_1 (x,k_{A\perp},\zeta_u) \bar f^\perp (y,k_{B\perp},\zeta_v) 
\nonumber\\ 
  && +  \frac{ \vec k_{A\perp}\cdot \vec q_\perp}{q_\perp^2} (f_1 
    - xf^{\perp})  (x, k_{A\perp},\zeta_u) \bar f_1 (y,k_{B\perp},\zeta_v) \biggr ].
\label{FAC}    
\end{eqnarray}
There is no contribution with $g^\perp$ or $\bar g^\perp$ in the symmetric part of $W^{\mu\nu}$. The result of $W^{-\mu}$ can be easily obtained 
through symmetries. 

\par 
The result in Eq.(\ref{FAC}) is derived in a relatively formal way. 
Its correctness needs to be examined, because possible singular contributions are hidden as we will show.
It is noted that the factorized result in Eq.(\ref{FAC}) also holds in the case that one replaces hadrons with partons, if the factorization is right. 
This gives a way to examine the result. For the case considered here, we can replace 
$h_A$ with a quark $q$ with the same momentum, and $h_B$ with an antiquark $\bar q$ with $P_B$. 
After the replacement, one can explicitly calculate TMD parton distributions of $q$ or $\bar q$ and the hadronic tensor 
$W^{+\nu}$. The obtained results can be used to check Eq.(\ref{FAC}).     
\par 
It is straightforward to calculate the quark-gluon correlator $\phi_3^\nu$ with $h_A=q$. At leading order of $\alpha_s$ we obtain 
\begin{eqnarray} 
\phi_3^\nu (x, k_{A\perp}) = \frac{\alpha_s C_F}{2\pi^2}\frac{k_{A\perp}^\nu}{k_{A\perp}^2} \frac{1}{1-x}
  + {\mathcal O}(\alpha_s^2). 
\label{T3T}          
\end{eqnarray} 
 The result is divergent at $x=1$. This clearly indicates the existence of light-cone singularity in the right-hand side of Eq.(\ref{W3G}). The left-hand side of Eq.(\ref{W3G}) does not have such a divergence. Therefore, the divergence can spoil the factorization if it is not correctly regularized. If we insert gauge links between operators in Eq.(\ref{PHI3}),  it will not generate extra contributions  at the considered order of $\alpha_s$. This indicates that the light-cone singularity is still there with the inserted gauge links off-light-cone. 
A solution to eliminate the singularity is to use the relation in Eq.(\ref{RE}) to express the factorization with $f_1$ and $f^\perp$ as given in Eq.(\ref{FAC}),   where TMD parton distributions are defined with the gauge links 
off light-cone in Eq.(\ref{DECMB}).   
But, as will be seen, the result is not exactly right.   
This can also be checked  with partonic states. 
\par 
\par 
\begin{figure}[hbt]
\begin{center}
\includegraphics[width=12cm]{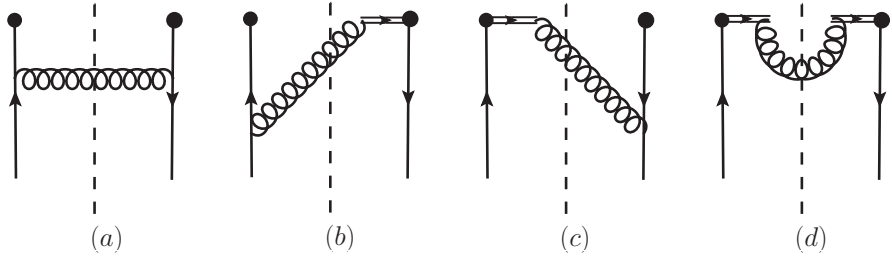}
\end{center}
\caption{The Feynman diagrams for the defined distributions. The double lines are for gauge links. The black dots denote the insertion of quark fields.    } 
\label{Feynman-dg2}
\end{figure}
\par
We first calculate $f^\perp (x,k_\perp)$ and $k_\perp^\mu f(x,k_\perp)$ by taking $h_A=q$. At leading order, 
the diagrams given in Fig.2 contribute. In Fig.2 the double lines represent the gauge link ${\mathcal L}_u$ 
or ${\mathcal L}_u^\dagger$. Calculating these diagrams we obtain:   
\begin{eqnarray} 
  f^\perp ( x, k_\perp, \zeta_u) &=& \frac{\alpha_s C_F}{2 \pi^2} \frac{1}{k_\perp^2} \biggr ( 
   -1 + \frac{1}{(1-x)_+} -\frac{1}{2} \delta (1-x) \ln\frac{k_\perp^2}{\zeta_u^2}\biggr ) + {\mathcal O}(\alpha^2_s),
\nonumber\\
k_\perp^\mu f_1 (x,k_\perp,\zeta_u) &=& \frac{\alpha_s C_F }{\pi^2}\frac{k_\perp^\mu}{k_\perp^2} 
\biggr [  \frac{x}{(1-x)_+} +\frac{1}{2}(1-x) - \frac{1}{2} 
  \delta (1-x)  \ln\frac{k_\perp^2}{\zeta_u^2}  -\frac{1}{2} \delta (1-x)  \biggr ] + {\mathcal O}(\alpha^2_s), 
\nonumber\\
  f_1 (x,k_\perp,\zeta_u) &=& \delta (1-x) \delta^2 (k_\perp)  + {\mathcal O}(\alpha_s),     
\label{T3TMD}  
\end{eqnarray} 
where we also list the leading order result of $f_1$. It is noted that the last term in $k_\perp^\mu f_1$ 
is from Fig.2d.  
The TMD antiquark distributions of an antiquark $\bar q (P_B)$ can be obtained as: 
\begin{equation} 
\bar  f_1 (x,k_\perp,\zeta_v) = f_1 (x,k_\perp,\zeta_v), \ \ \ \  \bar f^\perp ( x, k_\perp, \zeta_v) =
 f^\perp ( x, k_\perp, \zeta_v). 
\end{equation}  
If we take the gauge links in Eq.(\ref{DENMG}) along the direction $n^\mu$, the results of TMD quark distributions can be obtained by setting $\zeta_u =\infty$ in Eq.(\ref{T3TMD}). Then the results 
are divergent. This is the light-cone singularity. With the large but finite $\zeta_u$ the singularity 
is regularized. 
Using these results one can calculate the right-hand side of Eq.(\ref{FAC}). The result is:
\begin{eqnarray} 
 W^{+\nu}\biggr\vert_{\rm r.h.s\ of\ Eq.(\ref{FAC})} = \frac{q_\perp^\nu}{ P_B^-}  \frac{ \alpha_s C_F}{ 2 \pi^2 N_c  q_\perp^2}
  \biggr [  
    \frac{y\delta(1-x)}{(1-y)_+} + \frac{\delta(1-y)}{(1-x)_+} 
  + \delta(1-y)\delta (1-x) 
  \biggr (  \ln \frac{\zeta_v \zeta_u } {q^2_\perp}
     -1  \biggr ) \biggr ].
\label{WR}         
\end{eqnarray}
Now we calculate the left-hand side of Eq.(\ref{FAC}). With the replacement one needs to calculate $W^{+\nu}$ 
of the process $q(P_A) +\bar q(P_B) \to \gamma^* (q) +X$. At leading order $X$ is a gluon. The calculation is standard. We have after taking the limit $q_\perp \to 0$: 
\begin{equation}
  W^{+\nu} \biggr \vert_{\rm l.h.s.\ of\ Eq.(\ref{FAC})}   =\frac{q_\perp^\nu}{P_B^-}  \frac{ \alpha_s C_F}{ 2 \pi^2 N_c  q_\perp^2}
   \left [ \frac{ y \delta(1-x)}{(1-y)_+}
   + \frac{\delta(1-y)}{(1-x)_+} +\delta(1-x)\delta(1-y) \ln \frac{Q^2}{q^2_\perp} \right].
\label{WL}     
\end{equation}
Comparing Eq.(\ref{WR}) with Eq.(\ref{WL}) we find that the twist-3 TMD factorization as given in Eq.(\ref{FAC}) does not hold, even the light-cone singularities are 
regularized.  We notice that the difference between Eq.(\ref{WR}) and Eq.(\ref{WL}) appears in the region with $x=1$ and $y=1$, or it is proportional to $\delta(1-x)\delta(1-y)$. 
This indicates that the soft-gluon contribution in the process $q(P_A) +\bar q(P_B) \to \gamma^* (q) +g$ 
is not correctly factorized. The difference may be eliminated by taking $\zeta_u \zeta_v = e Q^2$. But one would like to have the factorization with arbitrary $\zeta_u$ and $\zeta_v$ for evolving TMD parton distributions at different energy scales. In our calculation of Fig.2 we notice that the contribution from Fig.2d 
to $k_\perp^\mu f_1$ is not zero and given by the last term in $k_\perp^\mu f_1$ of Eq.(\ref{T3TMD}).
This contribution gives the last term in Eq.(\ref{WR}).  
 But, 
the corresponding contribution from the process $q(P_A) +\bar q(P_B) \to \gamma^* (q) +X$ is absent.
We note that the exchanged gluon in Fig.2d is a soft gluon. 
It is clearly that this soft-gluon contribution 
should be subtracted. 
   
\par 
From the above discussion, the reason for the mismatch between Eq.(\ref{WR}) and Eq.(\ref{WL}) is that the soft-gluon contribution is not correctly factorized. This can be done correctly by introducing the soft 
factor: 
\begin{eqnarray}
S(\xi_\perp,\rho) = \frac{1}{N_c} \langle 0\vert {\rm Tr} \left  [   {\mathcal L}^\dagger_v (\xi_\perp)
  {\mathcal L}_u (\xi_\perp) {\mathcal L}_u^\dagger (\vec 0){\mathcal L}_v (\vec 0)   \right ] \vert 0\rangle,
\label{SoftS}
\end{eqnarray} 
with $\rho^2=(2 u\cdot v)^2/(u^2v^2)$. 
Using this soft factor we define the subtracted TMD parton distribution as:
\begin{eqnarray}       
f_1^{sub} (x,k_\perp)  = \int \frac{ d\xi^-d^2\xi_\perp } {2(2\pi)^3} e^{ -i x \xi^- P_A^+ - i \xi_\perp \cdot k_\perp} \frac{1}{\sqrt{S(\xi_\perp,\rho)}}
\langle h_A  \vert   \bar q(\xi) {\mathcal L}_u (\xi) \gamma^+ {\mathcal L}_u^\dagger (0)
  q (0)  \vert h_A  \rangle \biggr\vert_{\xi^+ =0}. 
\label{TMDSUB}   
\end{eqnarray} 
 Perturbatively, $S(\xi_\perp,\rho)= 1 +{\mathcal O}(\alpha_s)$. The perturbative result at one-loop with different regularizations
of I.R. divergences can be found in \cite{JMY,MWZ}. For our purpose here we only need the real contribution to $S$ at one-loop. The virtual part at one-loop does not depend on $\xi_\perp$.  The real contribution is given by the diagrams in Fig.2 with the quark lines replaced with the double lines
for the gauge links ${\mathcal L}_v$.  The result is:
\begin{equation} 
   S(\xi_\perp,\rho) = 1 -\frac{\alpha_s C_F} {2\pi^2 }\biggr ( 2 -\ln \frac{ (2 u\cdot v)^2}{u^2 v^2 } \biggr ) 
    \int d^2 k_\perp  e^{ik_\perp\cdot\xi_\perp} \frac{1}{k_\perp^2} + \cdots, 
\end{equation}     
where $\cdots$ stand for the contributions which are $\xi_\perp$-independent or at ${\mathcal O}(\alpha_s^2)$.
Using this result we have:
\begin{equation}  
k_\perp^\mu f_1^{sub} (x,k_\perp,\zeta_u) = \frac{\alpha_s C_F }{\pi^2}\frac{k_\perp^\mu}{k_\perp^2} 
\biggr [ \frac{x}{(1-x)_+} +\frac{1}{2}(1-x) - \frac{1}{2} 
  \delta (1-x) \biggr (  \ln\frac{k_\perp^2}{\zeta_u^2 } +\frac{1}{2}\ln\rho^2  \biggr ) \biggr ]  + {\mathcal O}(\alpha^2_s).  
\end{equation} 
We find that in $k_\perp^\mu f_1^{sub}$ the contribution from Fig.2d is subtracted.
Additional contribution related to regulators of light-cone singularities appears. This contribution will make 
the factorized $W^{+\nu}$ regulator-independent by noting $\rho^2 = \zeta_u^2 \zeta_v^2/(2 P_A^+ P_B^-)^2$.   
Similarly one can also define the subtracted TMD parton distribution $f^{\perp sub}$ with the factor 
$1/\sqrt{S(\xi_\perp,\rho)}$. The result of
$f^{\perp{sub}}$ of a quark at leading order $\alpha_s$ is the same as the unsubtracted.   
If we replace all TMD parton distributions in Eq.(\ref{FAC}) with the subtracted TMD parton distributions, 
instead of Eq.(\ref{WR}) the right-hand side of  Eq.(\ref{FAC}) is then given by 
\begin{equation}
  W^{+\nu} \biggr \vert_{\rm r.h.s\ of\ Eq.(\ref{FAC})}  =\frac{q_\perp^\nu}{P_B^-}  \frac{ \alpha_s C_F}{ 2 \pi^2 N_c  q_\perp^2}
   \left [ \frac{ y \delta(1-x)}{(1-y)_+}
   + \frac{\delta(1-y)}{(1-x)_+} +\delta(1-x)\delta(1-y) \ln \frac{Q^2}{q^2_\perp} \right]
\label{WRS}     
\end{equation}
for arbitrary $\zeta_u$ and $\zeta_v$. The above result exactly matches that in Eq.(\ref{WL}). 
Therefore, TMD factorization for $W^{+\nu}$ at the leading order of $\alpha_s$ only holds,  
provided that one uses the subtracted TMD parton distributions as the one defined in Eq.(\ref{TMDSUB}). 
In the result derived with the diagram expansion 
of Fig.1 the soft-gluon contribution is not correctly factorized, hence the derived factorization 
with unsubtracted TMD parton distributions is not correct. The correctly factorized form for $W^{+\nu}$ is still given 
by Eq.(\ref{FAC}), but with the subtracted TMD parton distributions.  For completeness we give the factorized form of $W^{-\nu}$:
\begin{eqnarray}   
W^{-\nu}  &=& \frac{q_\perp^\nu}{N_c P_A^+} \int d^2 k_{A\perp} d^2 k_{B\perp} \delta^2 (k_{A\perp}+k_{B\perp}-q_\perp)
  \biggr [  \frac{ \vec k_{A\perp}\cdot \vec q_\perp}{q_\perp^2} \bar f_1 (y,k_{B\perp},\zeta_v)  f^\perp (x,k_{A\perp},\zeta_u) 
\nonumber\\ 
  && +  \frac{ \vec k_{B\perp}\cdot \vec q_\perp}{q_\perp^2} (\bar f_1 
    - x \bar f^{\perp})  (y, k_{B\perp},\zeta_v)  f_1 (x,k_{A\perp},\zeta_u) \biggr ].
\label{FACN}
\end{eqnarray} 
In Eq.(\ref{FAC}, \ref{FACN}) all TMD parton distributions are subtracted ones. In the following, we will 
use the index $^{un}$ to denote unsubtracted TMD parton distributions. We add a factor $S^{-1/2}$ in the quark density matrix in Eq.(\ref{DENMG}) as in Eq.(\ref{TMDSUB}), so that the TMD parton distributions in Eq.(\ref{DECMB}) are subtracted ones.      
\par 
In the above we have studied the case in which the process is initiated by a quark from $h_A$ and an
antiquark from $h_B$. The factorization for the case with a quark from $h_B$ and an
antiquark from $h_A$ can be easily obtained from symmetries. Besides these cases, there can be those processes
where one gluon comes from $h_A$ or $h_B$ at leading order of $\alpha_s$. One can show that the contributions are included in Eq.(\ref{FAC},\ref{FACN}). This can also be shown explicitly by calculating $W^{+\nu}$ of the partonic processes with a gluon 
in the initial state and TMD quark distributions of the gluon. 
\par   
It is interesting to note that the twist-3 TMD factorization requires the soft factor at leading order of $\alpha_s$. This is in contrast to twist-2 TMD factorization for $W_\perp$ where the need of the soft factor can only be realized at next-to-leading order of $\alpha_s$. We also note that the way to implement the soft factor in twist-2- and twist-3 TMD factorization is different. To discuss this, we take $W_\perp$ and $f_1$ in $b$-space: 
\begin{equation} 
W_\perp ( x,y, b, Q) =\int d^2 q_\perp e^{-i\vec b \cdot \vec q_\perp} W_\perp (x,y,q_\perp, Q), \quad\quad 
f_1 (x,b,\zeta_u) =  \int d^2 k_\perp e^{-i\vec b \cdot \vec k_\perp}  f_1 (x,k_\perp, \zeta_u).
\end{equation} 
It can be shown that at leading power $W_\perp$ can be factorized as\cite{CSS,JMYP}: 
\begin{equation} 
   W_\perp(x,y,b, Q) =  {\mathcal H} (\zeta_u,\zeta_v,\rho, Q) \frac{1}{S(b, \rho)} f_1^{un}(x,b,\zeta_u) \bar f_1^{un}(y, b,\zeta_v)
\label{WTH} 
\end{equation} 
with the perturbative coefficient ${\mathcal H} =1 +{\mathcal O}(\alpha_s)$. Now one has certain freedom 
to define subtracted TMD parton distributions. E.g., in \cite{JMY} one defines the subtracted TMD parton distributions by dividing the unsubtracted one by $S$, so that $W_\perp$ takes the form 
$W_\perp ={\mathcal H} f_1 \bar f_1 S$. 
One can also define the subtracted TMD parton distributions by dividing the unsubtracted one with $\sqrt{S}$. 
Then $W_\perp$ takes the form  
$W_\perp ={\mathcal H} f_1 \bar f_1$. Different versions of subtracted TMD parton distributions exist\cite{JC1,LTMD}.   But in twist-3 factorization for $W^{\pm\nu}$, one has to define the subtracted TMD parton distributions  as the unsubtracted ones divided by $\sqrt{S}$
in order to make the factorization correctly at leading order of $\alpha_s$. 
The reason for this is that $f_1(x, k_\perp)$ appears in our case always in combination with $k_\perp^\mu$. The one-loop virtual part 
of the soft factor will not contribute to $k_\perp^\mu f_1(x,k_\perp)$ at the considered order. Taking the square root of $S$ 
the unwanted contribution from Fig.2d is eliminated. If we write the factorized form of $W^{\pm\nu}$ with the soft factor explicitly, as we can do this in the case of twist-2, then the virtual part of $S$ may contribute at the considered order. This will spoil the factorization, since there is no  exchange 
of virtual gluon at the considered order.

\par 
It should be noted that light-cone singularities can be regularized in different ways. This introduces a scheme-dependence in 
TMD factorization. TMD parton distributions can be defined differently in different schemes. In this letter, we take the gauge-links off light-cone in the unsubtracted TMD parton distributions to regularize 
the singularity. In \cite{JC1} the gauge links in the unsubtracted TMD parton distributions are along light-cone directions, the light-cone singularities are subtracted by a soft factor, which is defined with gauge links along-light-cone directions and those off  light-cone. The difference between the two schemes results in that the perturbative coefficient ${\mathcal H}$
in Eq.(\ref{WTH}) is different. This has been discussed in detail in \cite{PSFY}.  
TMD factorization at leading twist has been studied in 
the framework of SCET in \cite{EIC1, EIC2, CJNR}. In SCET the light-cone singularities are  regularized differently. In \cite{EIC1, EIC2} all relevant gauge links are along light-cone. The related singularities are regularized with the so-called $\Delta$-regulators. The twist-2 TMD factorization in this scheme is equivalent to that
with TMD parton distributions in \cite{JC1} as shown in \cite{EIC2,JCTR}. 
In the scheme with $\Delta$-regulators it is found in \cite{EIC1} that the subtracted TMD distribution $f_1$  
has to be defined as the unsubtracted $f_1^{un}$ divided by the square root of the soft factor in order to have a clear 
separation of different momentum modes. This is consistent with ours although the reasons are different.  
In \cite{CJNR} the so-called $\nu$-regulator is employed. Twist-2 TMD parton distributions of an unpolarized hadron   
has been studied in SCET already at two-loop level in \cite{GLY,ESV} and the relevant soft factor has been calculated 
at two-loop\cite{ESVSoft} and at three-loop\cite{LZ,AV}.

\par 
As discussed in the above, TMD factorization at twist-2 has been studied intensively. In this letter we have only studied 
twist-3 TMD factorization at leading order of $\alpha_s$. At this order we already meet light-cone singularities and the problem of how to factorize the soft-gluon effect. Our factorized results in Eq.(\ref{FAC}, \ref{FACN}) hold at the leading order of $\alpha_s$. Beyond this order, the factorized form may be changed, because some additional effects 
of soft gluons can appear and are not included in the factorized form derived at the leading order. To clarify this, 
one needs to examine the factorization at higher orders and eventually to prove the factorization at all orders. It is noted that in Eq.(\ref{FAC}) we have in fact two perturbative coefficients because there are  two different  
combinations of TMD parton distributions. The two perturbative coefficients are $1$ at the leading order and will 
receive corrections from higher orders of $\alpha_s$. Beyond the leading order the coefficients will 
depend on light-cone regulators $\zeta_{u,v}$ and the renormalization scale $\mu$. These dependences will 
be cancelled by those of TMD parton distributions. The $\mu$-dependences of unsubtracted 
TMD parton distributions are simply determined by the anomalous dimension of the quark field in the axial gauge.  
The $\zeta_u$-dependence of TMD parton distributions is an interesting aspect of TMD factorization, which we will discuss 
in the below. 
\par 
Unlike parton distributions in collinear factorization, which do not depend on the energy of the hadron, 
TMD parton distributions depend on the energy of the hadron through the variable $\zeta_{u,v}$. The dependence is determined by Collins-Soper equation. Starting with TMD factorization of $W_\perp$ and using the equation for $f_1$ one can resum contributions with the large logarithms of $q_\perp/Q$ in the perturbative coefficient functions in collinear factorization 
of $W_T$\cite{CSS}. Large logarithms of $q_\perp/Q$ in $W_\Delta$ may also be re-summed with TMD factorization 
of $W^{\pm\mu}$. For this one needs the Collins-Soper equation of $f_1$ and $f^\perp$. The latter is unknown and will be derived here. 
\par 
Collins-Soper equations are conveniently expressed in $b$-space. The definition of $f^\perp$ in $b$-space 
is
\begin{eqnarray} 
-i \partial_b^\mu  f^\perp (x, \xi_\perp,\zeta_u) &=&P_A^+  \int \frac{ d\xi^- } {4\pi} e^{ -i x \xi^- P_A^+ }\frac{1}{\sqrt{S(\vec \xi_\perp, \rho)}}
\langle h_A  \vert   \bar q(\xi) {\mathcal L}_u (\xi) \gamma_\perp^\mu{\mathcal L}_u^\dagger (0)
  q (0)  \vert h_A  \rangle \biggr\vert_{\xi^+=0}
\end{eqnarray}
with $\vec b =\vec\xi_\perp$ and $\partial_b^\mu =\partial/\partial b_\mu$. The dependence on $P_A^+$ is determined by the same dependence of the unsubstracted $f^{\perp un}$. Since the dependence is through the variable $\zeta_u$, we need to calculate 
the derivative of $f^{\perp un}$ with $\zeta_u$. The detail about how to calculate the dependence has been  
discussed in \cite{CS,CSEQ1}. We will not discuss it here. 
\par 

\begin{figure}[hbt]
\begin{center}
\includegraphics[width=17cm]{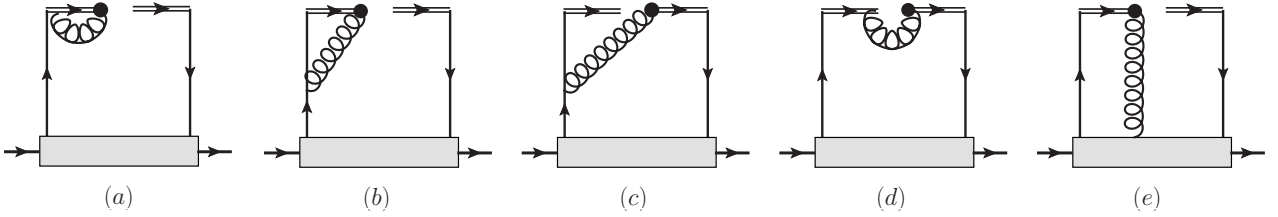}
\end{center}
\caption{The Feynman diagrams for the $\zeta_u$-evolution. The black dots denote the derivative acting on 
the gluon attached to the gauge links. The gray boxes stand for quark density matrix defined before.    } 
\label{Feynman-EQ}
\end{figure}

\par 

At leading order the derivative receives the contributions from Fig.3. In diagrams of Fig.3 the black dots 
denote the derivative acting on the vertex: 
 \begin{equation}   
   \zeta_u \frac {\partial }{\partial \zeta_u}       =  \tilde u^\mu \frac{\partial }{\partial u^\mu} 
\end{equation}  
with  $\tilde u^\mu = (- u^+, u^-,0,0)$. Not all diagrams give nonzero contributions. Fig.3a and Fig.3d 
give no contribution. The contribution from Fig.3e is power-suppressed and can be neglected.
The contributions from Fig.3b or Fig.3c contains I.R. divergence, the divergence is canceled at the end. 
From Fig.3 we derive the Collins-Soper equation for $f^\perp$: 
\begin{eqnarray} 
   \zeta_u \frac{\partial}{\partial \zeta_u} \left (  \partial_b^\mu  f^{\perp}(x,b,\zeta_u)  \right ) &=& 
      \frac{\alpha_s C_F}{2\pi x}\biggr [  - \ln\frac{x^2 b^2 \zeta_u^2 e^{2\gamma }}{4}
      \partial_b^\mu \left ( x  f^\perp (x, b,\zeta_u)  + f_1 (x,b, \zeta_u)  \right )  
\nonumber\\      
     &&   + 2 \frac{b^\mu} {b^2} f_1 (x,b,\zeta_u )       \biggr ] + {\mathcal O}(\alpha_s^2) .  
\label{CSP}           
\end{eqnarray} 
Comparing the $\zeta_u$-evolution of $f_1$ calculated in \cite{CS,CSS}:  
\begin{eqnarray} 
 \zeta_u \frac{\partial}{\partial \zeta_u}  f_1 (x,b,\zeta_u,\mu) 
 = -\frac{\alpha_s C_F}{\pi} \ln \left (  \frac{ x^2\zeta_u^2 b^2}{4} e^{2\gamma-1}\right ) f_1(x,b,\zeta_u,\mu)  + {\mathcal O}(\alpha_s^2), 
\label{CSf1}    
\end{eqnarray}
we note that the $\zeta_u$-evolution of $f^\perp$ is mixed with $f_1$, or Collins-Soper equation of $f^\perp$ 
is not homogeneous in itself. It is noted that the log term in the above evolutions can be written in the $K+G$ form\cite{CS, CSS}:
\begin{equation}  
   K(\mu,b) + G(\mu,x\zeta_u) = -\frac{\alpha_s C_F}{\pi} \ln \left (  \frac{ x^2\zeta_u^2 b^2}{4} e^{2\gamma-1}\right ) + {\mathcal O}(\alpha_s^2),  
\end{equation} 
where $K$ is the soft gluon contribution and $G$ is the hard-gluon contribution. 
The sum $K+G$ does not depend on $\mu$. The $\mu$-dependence of $K$ is given by:
\begin{equation} 
  \mu \frac {\partial K (\mu,b)}{\partial \mu} = -\gamma_K, \quad \gamma_K =2 \frac{\alpha_s C_F}{\pi} +{\mathcal O}(\alpha_s)
\end{equation} 
where $\gamma_K$ is the cusp anomalous dimension\cite{CUSP}. In the case of twist-2 TMD parton distributions $K(\mu,b)$ can also be determined by the 
soft factor\cite{CS, CSS}.          
\par 
Since Collins-Soper equation for $f^\perp$ is nonhomogeneous, the resummation of large logarithms of $q_\perp/Q$ in collinear factorization of $W_\Delta$ or $W_{l,n}$ will be complicated.  We note here that the  resummed form of $W_\perp$ is relatively simple, because Collins-Super equation for $f_1$ is homogeneous. This enables 
to get the resummed form of $W_\perp $ at high $Q^2=Q_H^2$ as the product of $W_\perp$ at a low $Q^2=Q_L^2$ multiplied with 
Sudakov factor. By taking $Q_L \sim 1/b$, $W_\perp(x,y,b,Q_L)$ does not contain large logarithms terms. 
The large logarithms terms in $W_\perp(x,y,b,Q_H)$ are resummed in the Sudakov factor\cite{CSS,JMY}.
But for $W_\Delta$ or $W_{l,n}$ the resummed form will be not so simple as that of $W_T$.
We leave this for a future study. 
\par 
To summarize: Some structure functions in Drell-Yan processes can only be factorized 
with twist-3 TMD parton distributions. We have studied twist-3 TMD factorization of these structure functions.  Although the twist-3 factorization 
can be derived at leading order of $\alpha_s$ with diagram expansion in a straightforward way, but light-cone singularities exist already at this order. We use off light-cone gauge links to regularize the singularities. With partonic results we have shown that even at leading order of $\alpha_s$ one needs 
to introduce a soft factor to correctly factorize soft-gluon contributions. The correct TMD factorization at twist-3 
is obtained only with the subtracted TMD parton distributions. We have derived Collins-Soper equation for one twist-3 TMD parton distribution relevant here and found that the equation is nonhomogeneous. This will result in that the resummation of large logarithms term in the corresponding structure function is complicated. In this work, we have only studied the factorization with chirality-even operators. In the future we will study the case with chirality-odd operators
and the resummation.

\par\vskip20pt      

\noindent
{\bf Acknowledgments}
\par
We thank Dr. J.W. Qiu for interesting suggestion and discussions. 
The work is supported by National Nature
Science Foundation of P.R. China(No.11275244, 11675241). The partial support from the CAS center for excellence in particle 
physics(CCEPP) is acknowledged. 
 
\par\vskip40pt

\par

\end{document}